\documentclass[journal,10pt,twocolumn]{IEEEtran}
\usepackage[table]{xcolor}
\usepackage{comment}
\usepackage{multicol}
\usepackage{color, colortbl}
\usepackage{amsmath,amssymb,amsthm}

\usepackage{cite}
\usepackage{pdflscape}
\usepackage[english]{babel}
\usepackage{multicol}
\usepackage{multirow}
\usepackage{algorithm}
\usepackage[noend]{algpseudocode}

\usepackage{amsmath}
\usepackage{amsmath}
\usepackage{amsfonts}
\usepackage{amssymb}
\usepackage{float}
\usepackage{flushend}
\usepackage{stfloats}
\usepackage{subfigure}
\usepackage{float}
\usepackage{amsmath}
\usepackage{mathrsfs}
\usepackage{color}
\usepackage{float}
\usepackage{array}
\usepackage{balance}
\usepackage{graphicx}
\usepackage{epsfig}
\usepackage{epstopdf}
\usepackage{textcomp}
\usepackage{url}
\usepackage{footnote}
\usepackage{comment}
\usepackage[numbers, square, comma, sort&compress]{natbib}
\usepackage[numbers]{natbib}

\usepackage[singlespacing]{setspace}
\usepackage{etoolbox}
\usepackage{ccaption}
\usepackage[font=footnotesize,labelfont=rm,figurename=Fig.]{caption}
\IEEEoverridecommandlockouts

\newcounter{inlineequation}
\DeclareMathAlphabet{\mathpzc}{OT1}{pzc}{m}{it}
\definecolor{Gray}{gray}{0.9}
\definecolor{LightCyan}{rgb}{0.88,1,1}

\begin{document}

\title{Coverage gain and Device-to-Device user Density: Stochastic Geometry Modeling and Analysis}

\author{Hafiz Attaul Mustafa, Muhammad Zeeshan Shakir, Muhammad Ali Imran, Ali Imran, and Rahim Tafazolli
\fontsize{10}{10}\selectfont

\thanks{Manuscript received x x, 2015; revised x x, 2015; accepted x x, 2015. Date of publication x x, 2015; date of current version x x, 2015. This work was supported by Qatar National Research Fund (a member of the Qatar Foundation, NPRP grant No. 5-1047-2437) and members of the University of Surrey 5GIC (http://www.surrey.ac.uk/5gic). The associate editor coordinating the review of this paper and approving it for publication was J. Lee.}

\thanks{H. A. Mustafa, M. A. Imran, and R. Tafazolli are with Institute for Communication Systems (ICS), University of Surrey, Guildford, UK. Emails: \{h.mustafa, m.imran, r.tafazolli\}@surrey.ac.uk}

\thanks{M. Z. Shakir is with Electrical and Computer Engineering Dept., Texas A\&M University at Qatar,  Doha. Email: muhammad.shakir@qatar.tamu.edu}

\thanks{A. Imran is with School of Electrical and Computer Engineering, University of Oklahoma, Tulsa, USA. Email: ali.imran@ou.edu}

}

\maketitle
\begin{abstract}
Device-to-device (D2D) communication has huge potential for capacity and coverage enhancements for next generation cellular networks. The number of potential nodes for D2D communication is an important parameter that directly impacts the system capacity. In this paper, we derive analytic expression for average coverage probability of cellular user and corresponding number of potential D2D users. In this context, mature framework of stochastic geometry and Poisson point process has been used. The retention probability has been incorporated in Laplace functional to capture reduced path-loss and shortest distance criterion based D2D pairing. The numerical results show a close match between analytic expression and simulation setup.
\end{abstract}


\section{Introduction}\label{intro}
\IEEEPARstart{T}{he} homogeneous Poisson point process (PPP) and its variant marked PPP (MPPP) has special characteristics to provide mathematical tractability for modeling the spatial distribution of macro cells, small cells, and cellular users. These tools are used in \cite{xu_transmission_2013, 7056528, 6953066} to derive the analytic expression for transmission capacity and outage probability of cellular and D2D users; however, no notion of D2D pairing, based on some selection criterion (e.g., reduced path-loss), has been assumed in Laplace functional of point process. This means every node is in D2D communication even if the nodes are not in  a feasible cooperation region\footnote{The distance between nodes can be as large as twice the radius of the cell which is not a realistic assumption for direct communication.}. The maximum achievable transmission capacity of D2D communication in heterogeneous networks with multi-bands has been analyzed in \cite{6504260}. The problem has been formulated as a sum capacity optimization problem for D2D network where outage probabilities of cellular and D2D users are set as constraints. The results of \cite{6504260} are based on MPPP for D2D user density, however, similar to the previous papers, no analytic representation of some criterion-based D2D pairing has been considered.

In this letter, we present an analytic framework for the analysis of average coverage probability of cellular user and corresponding D2D user density. Using mature framework of stochastic geometry and MPPP, we introduce retention probability in Laplace functional of MPPP to capture the effect of reduced path-loss based selection of D2D pairs. By assuming every node in D2D communication (no D2D pairing criterion) and using same transmit power for D2D pairs, a lower bound on average coverage probability has been introduced as a special case.
  
The system model is presented in Section II. Section III covers retention probability followed by derivation of closed-form expression for average coverage probability of cellular user. Numerical results are presented in Section IV followed by conclusion in Section V.
\section{System Model}\label{sys_mod}
\begin{figure}[!htb]
\centering
\includegraphics[width = 0.95\columnwidth, height = 1.75in]{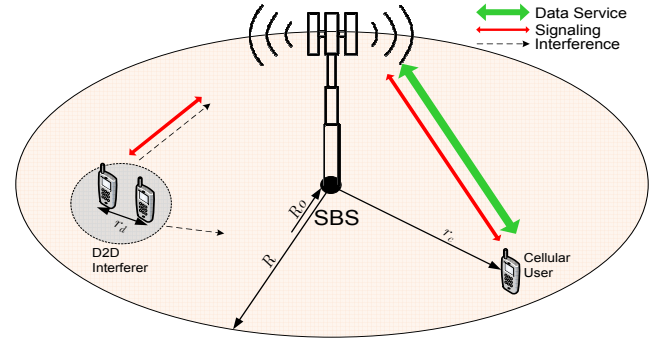}
\caption{System model for D2D user density.}\label{Figure:sm}
\vspace{-2mm}
\end{figure}

Consider small cell base station (SBS), cellular user, and potential D2D users as shown in Fig. \ref{Figure:sm}. 
In this model, the uplink resources of cellular users are shared by potential D2D users. The time division duplex (TDD) mode is assumed between D2D users. The data and signaling is provided by the SBS to the cellular users whereas only signaling is assumed for potential D2D users. For average coverage probability of cellular user, interference is generated by all D2D pairs.
 
The cellular and potential D2D users are distributed in the coverage area bounded between SBS radius $R$ and the protection region $R_0$. All potential D2D users can make pair with each other, however, due to the small probability of occurrence, pairing of more than two D2D users with the reference user is ignored in this work. The distance between SBS and cellular user is $r_c$ which is used to calculate path-loss. Every successful D2D pair has a distance of $r_d$ between nodes. The channel model assumes distance dependent path-loss and Rayleigh fading. The simple singular path-loss model $({r_c}^{-\alpha})$ is assumed where the protection region ensures the convergence of the model by avoiding $r_c$ to lie at the origin. However, $R_0 \ll R$ such that it can be considered as an atom in point process terminology i.e., $R_0 \sim 0$. The received power at SBS follows exponential distribution. The distance $r_c$ follows uniform probability distribution function (pdf) as follows:
\begin{align}
f(r_c)=\frac{2r_c}{R^2},f({\theta})=\frac{1}{2\pi},
\label{frc}
\end{align}
where $R_0\leq r_c \leq R$ and $0 \leq \theta \leq 2\pi$.
\begin{figure}[t]
\centering
\includegraphics[width=1\columnwidth]{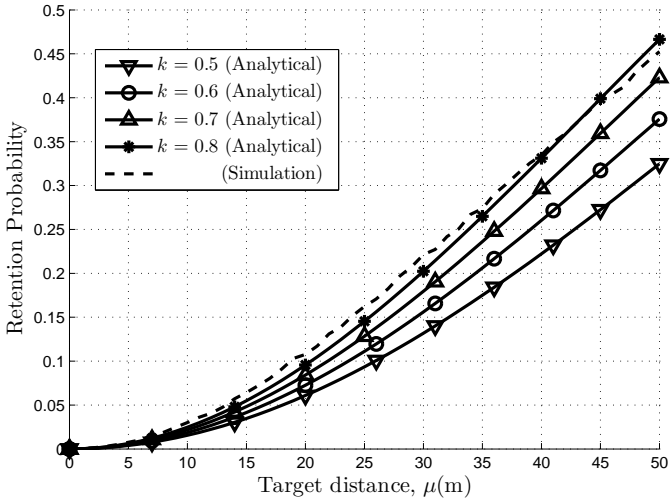}
\caption{Retention probability and tuning factor $k$ for $\lambda = 0.000025$.}\label{Figure:k_analysis}
\vspace{-2mm}
\end{figure}
\section{Retention and Coverage Probability}\label{retprob}
We assume homogeneous MPPP for distribution of potential D2D users to analyze the average coverage probability of cellular user. The homogeneous MPPP $\Phi = \{x_{i},y_{i}\}$ is defined by marks and labels. The marks represent the spatial positions on the plane $x_{i} \in \mathbb{R}^2$ and labels represent the channel gains $y_{i} \in \mathbb{R}_+$ of D2D users. In order to incorporate selection criterion for D2D pairing, we assume reduced path-loss (shortest distance) between potential D2D users. All nodes that do not meet this criterion are excluded from D2D pairing. In this context, we convert $\Phi$ to $\Phi^p$ where $p: \mathbb{R}^2 \mapsto [0, 1]$ performs thinning process to analytically capture the shortest distance based selection of points.

The number of points that generate shot-noise field (SN) at a reference point are implicitly represented in Laplace functional of MPPP  \cite{blaszczyszyn_stochastic_2009} given as:
\begin{align}
\mathcal{L}_{\Phi} (f) = & \,e^{- \int_{\mathbb{R}^d} (1-e^{-f(x)}) \lambda \, dx},
\label{RetProb}
\end{align}
where $f(x)$ is the real function defined on $\mathbb{R}^d$ and $\lambda$ is the intensity measure of points (in this case D2D user density). 
If we introduce retention probability $p(r_d<\mu)$ as the probability that certain points meet some target distance $\mu$, then we can incorporate it in Laplace functional of MPPP. For notational convenience, we will use $p(r_d)$ in Laplace functional (\ref{RetProb}) as:
\begin{align}
\mathcal{L}_{\Phi^p} (f) = & \,e^{- \int_{\mathbb{R}^d} (1-e^{-f(x)}) p(r_d) \lambda \, dx},
\label{RetProb_1}
\end{align}
In this paper, we introduce $p(r_d)$ (based on \cite{6909030}) with tuning factor $k$ to analytically capture simple reduced path-loss (shortest distance) based selection of D2D pairs as follows:
\begin{align}
p(r_d) = & \, 1-e^{-k \pi \lambda \mu^{2}},
\label{prd}
\end{align}

\begin{figure}[t]
\centering
\includegraphics[width = 1\columnwidth]{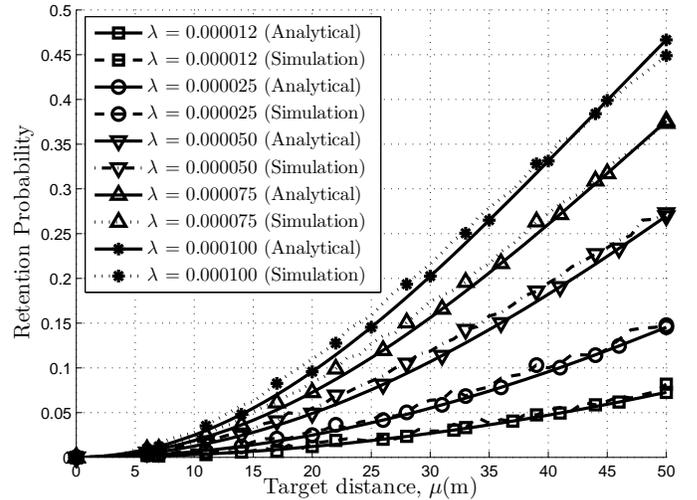}
\caption{Retention probability for $k = 0.8$, and different $\lambda$.}\label{Figure:k_lambda_analysis}
\vspace{-2mm}
\end{figure}

The tuning factor has been determined by simulating the system model. The retention probability has been plotted for D2D user density $\lambda$, target distance $\mu$, and different values of $k$. For illustrative purpose, the value of $\lambda$ = 0.000025 has been chosen as shown in Fig. \ref{Figure:k_analysis}. The value of $k=$ 0.8 matches with the simulation runs for lower values of target distance (e.g., $\mu \leq$ 50m). For this value of $k$, the retention probability for different values of $\lambda$ has been plotted in Fig. \ref{Figure:k_lambda_analysis}. In this figure, retention probability depends on two parameters i.e., $\lambda$ and $\mu$. The higher value of $\lambda$ results in larger number of potential D2D pairs within a target distance; however, shortest distance criterion allows only single pair for D2D communication. This results in mismatch between analytic expression and simulation curves. For example, the value of $\lambda =$ [0.000050, 0.000075, 0.0001] results in higher mismatch as compared to the lower values ([0.000012, 0.000025]). This mismatch can be removed if point-to-multipoint D2D links are considered for direct communication. The higher $\lambda$ also increases the retention probability which is intuitive as more D2D links can meet the target  distance criterion. The target value $\mu$ has similar interpretation. For lower target values, few D2D interferers lie in target distance. This results in close match between analytic expression of retention probability and simulation curves. The higher target distance results in more interferers which is not instructive for realistic scenarios.

In order to derive average coverage probability of cellular user, we assume interference-limited environment ($\sigma^2 = 0$). Hence, the signal-to-interference ratio (SIR) is given as:
\begin{align}
\textnormal{SIR}_{SBS} =  \frac{p_c f_{c} r^{-\alpha}_{c}}{\sum_{i \in \Phi} p_{i} f_{i} r^{-\alpha}_{i}},
\label{SIR_S1}
\end{align}
where $p_{c}$, and $p_{i}$ are the transmit power of cellular user and D2D interferers, respectively; $f_{c}$, and $f_{i}$ are respective small-scale fading. The corresponding distance dependent path-losses are $r^{-\alpha}_{c}$, and $r^{-\alpha}_{i}$.

The average coverage probability of cellular user distributed uniformly over plane between $R$ and $R_{0}$ at a distance $r_{c}$ from the serving SBS is given as:
\begin{align}
p_{cov}^{c} = & \,\mathbb{E}_{r_{c}}\big[\mathbb{P}[\textnormal{SIR}_{SBS} \geq \gamma]\,|\,r_{c}\big], \IEEEnonumber
\\ =& \,\mathbb{E}_{r_{c}}\big[\mathbb{P}[(f_{c} \geq \frac{\gamma I_{m}}{p_{c} r^{-\alpha}_{c}})]\,|\,r_{c}\big],	\label{SIR_AppA}
\end{align}
where 
\begin{align}
I_{m} = \sum_{i \in \Phi} p_{i} f_{i} r^{-\alpha}_{i},
\label{Im}
\end{align}
is the interference due to D2D users in the coverage area.

In (\ref{SIR_AppA}), the coverage probability depends on a number of random variables e.g., $p_{c}, f_{c}, r^{-\alpha}_{c}, p_{i}, f_{i}, r^{-\alpha}_{i}$. The power transmitted by the cellular user $p_{c}$ is assumed to be independent of the interferers and controlled by the serving SBS. The fading $f_{c}$ and $f_i$ follows Rayleigh distribution with $p_c$ and $p_i$ as exponentially distributed. The cellular user is uniformly distributed in the coverage area of SBS whereas all interferers are distributed according to MPPP process.
Conditioning on $g = \{p_{i}, f_{i}\}$, the coverage probability of cellular user for a given transmit power $p_c$ is
\begin{align}
\mathbb{P}[\textnormal{SIR}_{SBS} \geq \gamma]\,|\,r_{c},g =& \, \int_{x=\frac{\gamma I_{m}}{p_{c} r^{-\alpha}_{c}}}^\infty e^{- x}dx,	\IEEEnonumber
\\ =& \, e^{-\gamma p^{-1}_{c} r^{\alpha}_{c} I_{m}},
\label{SIR_AppA1}
\end{align}
De-conditioning by $g$, (\ref{SIR_AppA1}) results into:
\begin{align}
\mathbb{P}[\textnormal{SIR}_{SBS} \geq \gamma]\,|\,r_{c} =& \, \mathbb{E}_{g}\big[e^{-\gamma p^{-1}_{c} r^{\alpha}_{c} I_{m}}\big],	\IEEEnonumber
\\ =& \, \mathbb{E}_{g}\big[e^{- s_{c} I_{m}}\big],	\IEEEnonumber
\\ =& \,\mathcal{L}_{I_{m}}\big(s_{c}\big),
\label{SIR_AppA2}
\end{align}
where $s_{c} = \gamma p^{-1}_{c} r^{\alpha}_{c}$. 
\\
Putting the value of $I_{m}$ from (\ref{Im}) into (\ref{SIR_AppA2})
\begin{align}
\mathcal{L}_{I_{m}}\big(s_{c}\big) = & \,\mathbb{E}_{\Phi,p_{i},f_{i}}\bigg[ e^{-s_{c} \sum_{i \in \Phi} p_{i} f_{i} r^{-\alpha}_{i}}\bigg],	\IEEEnonumber
\\  = & \,\mathbb{E}_{\Phi,p_{i},f_{i}}\bigg[\prod_{i \in \Phi} e^{-s_{c}  p_{i} f_{i} r^{-\alpha}_{i}}\bigg],	\IEEEnonumber
\\ = & \,\mathbb{E}_{\Phi,p_{i}}\bigg[\prod_{i \in \Phi} \mathbb{E}_{f_{i}} \big[e^{-s_{c}  p_{i} f_i r^{-\alpha}_{i}}\big]\bigg],	\IEEEnonumber
\\ = & \,\mathbb{E}_{\Phi}\bigg[\prod_{i \in \Phi} \mathbb{E}_{p_{i}} \bigg(\frac{1}{1+s_{c}  p_i r^{-\alpha}_{i}}\bigg)\bigg],	\IEEEnonumber
\\ = & \,\mathbb{E}_{\Phi}\bigg[\prod_{i \in \Phi} \underbrace{\bigg(\frac{1}{1+s_{c} \mathbb{E} [p_i] r^{-\alpha}_{i}}\bigg)}_{f(x)}\bigg],
\label{SIR_AppA3}
\end{align}
where (\ref{SIR_AppA3}) results from the i.i.d distributions of $p_{i}$ and $f_{i}$ and further independence from the underlay MPPP process.

The probability generating functional (PGFL) for a function $f(x)$ with retention probability $p(r_d)$ from (\ref{prd}) implies:
\begin{align}
\mathbb{E}\bigg[\prod_{m \in \Phi}f(x) \bigg] = & \,e^{- \int_{\mathbb{R}^2} (1-f(x)) p(r_d) \lambda \, dx},	\IEEEnonumber
\\ = & \,e^{-\lambda\, p(r_d) \int^{\infty}_{0} \int^{2\pi}_{0} (1-f(x)) x dr dx},	\IEEEnonumber
\\ = & \,e^{-2\pi \lambda\, p(r_d) \int^{\infty}_{0} (1-f(x)) x dx},
\label{PGFL_AppA}
\end{align}
Putting $f(x)$ from (\ref{SIR_AppA3}) into (\ref{PGFL_AppA}) results:
\begin{align}
\mathcal{L}_{I_{m}}\big(s_{c}\big) = & \,\,e^{-2\pi \lambda\, p(r_d) \int^{\infty}_{R_0} \big(1- \frac{1}{1+s_{c}\mathbb{E}[p_{i}] x^{-\alpha}} \big) x dx},	\IEEEnonumber
\\ = & \,\,e^{-2\pi \lambda\, p(r_d) \int^{\infty}_{R_0} \big(\frac{1}{1+\frac{ x^{\alpha}}{s_{c}\mathbb{E}[p_{i}]}} \big) x dx},
\label{PGFL_AppA1}
\end{align}
By substituting $\frac{x^{\alpha}}{s_{c}\mathbb{E}[p_{i}]} = u^{\alpha}$, (\ref{PGFL_AppA1}) results
\begin{align}
\mathcal{L}_{I_{m}}\big(s_{c}\big) = & \,\,e^{-2\pi \lambda \, p(r_d) (s_{c})^{\frac{2}{\alpha}} \mathbb{E}[{p^{\frac{2}{\alpha}}_{i}}] \int^{\infty}_{R_0} \big(\frac{u}{1+u^{\alpha}}\big) du},
\label{PGFL_AppA2}
\end{align}
Since $R_0 \ll R$, therefore assuming $R_0 \sim 0$, the integral on right hand side of (\ref{PGFL_AppA2}) can be evaluated as:
\begin{align}
\int^{\infty}_{0} \bigg(\frac{u}{1+u^{\alpha}}\bigg) du = & \frac{\pi }{\alpha \sin(\frac{2 \pi}{\alpha})},
\label{PGFL_AppA3}
\end{align}
Putting (\ref{PGFL_AppA3}) into (\ref{PGFL_AppA2}) and using uniform distribution from (\ref{frc}), the average coverage probability of a cellular user (\ref{SIR_AppA}) is:
\begin{align}
p_{cov}^{c} = & \,\mathbb{E}_{r_{c}}\bigg[e^{-\frac{2\pi^2 \lambda \, p(r_d) \,r^{2}_{c}}{\alpha \sin(\frac{2 \pi}{\alpha})} \big(\frac{\gamma}{ p_{c}}\big)^{\frac{2}{\alpha}} \mathbb{E}[{p^{\frac{2}{\alpha}}_{i}}]}\,|r_{c}\bigg], \IEEEnonumber
\\ =& \int^{R}_{R0}e^{-\frac{2\pi^2 \lambda (1-e^{-k \pi \lambda \mu^2})\,r^{2}_{c}}{\alpha \sin(\frac{2 \pi}{\alpha})} \big(\frac{\gamma}{ p_{c}}\big)^{\frac{2}{\alpha}} \mathbb{E}[{p^{\frac{2}{\alpha}}_{i}}] }\,\frac{2 r_{c}}{R^{2}} dr_{c},
\label{SIR_AppA5}
\end{align}

For same transmit power of all D2D interferers, the average coverage probability of cellular user for path-loss exponent $\alpha = 4$ and $R_0 \sim 0$ reduces to:
\begin{align}
p_{cov}^{c} = & \,\, \frac{e^{-\frac{\pi^2 R^2 \lambda}{2} \sqrt{\frac{\gamma p_i}{p_c}}(1-e^{-k \pi \lambda \mu^2})}-1}{-\frac{\pi^2 R^2 \lambda}{2}\sqrt{\frac{\gamma p_i}{p_c}}(1-e^{-k \pi \lambda \mu^2})},
\label{PGFL_SC3}
\end{align}

The lower bound on average coverage probability of cellular user is simply the cellular coverage probability without PPP thinning process. It can be derived by relaxing the shortest distance criterion and allowing every node to be in D2D pair. This case shows maximum interference due to full frequency reuse by all nodes in the coverage area. If we assume $p(r_d) = 1$, it means target distance $\mu$ in (\ref{prd}) has no distance constraint and all nodes in the coverage area can communicate directly on same channel as used by the cellular user. In this case, the lower bound on average coverage probability of cellular user is given as:
\begin{align}
p_{cov, lb}^{c} = & \frac{e^{\frac{-\pi^2 R^2 \lambda}{2} \sqrt{\frac{\gamma p_i}{p_c}}}-1}{\frac{- \pi^2 R^2 \lambda}{2}\sqrt{\frac{\gamma p_i}{p_c}}},
\label{corollary1Equ}
\end{align}
\section{Numerical Results}
In this section, we numerically evaluate the analytic expressions of Sec. \ref{retprob} by varying the number of different parameters for a fixed cell of radius $R$ = 500m and a radius of protection region $R_0$ = 1m. The cell-edge effect is simulated by dropping D2D users around cell boundary. The D2D pairs are chosen on reduced path-loss (shortest distance) criterion which is captured analytically by thinning the Poisson point process using retention probability (\ref{prd}). The power ratio of cellular user and D2D transmitter is assumed to be 500 with $p_c$ = 100mW and $p_i$ = 0.2mW. In order to calculate the average effect of coverage probability, 3000 monte-carlo simulations are run.

\begin{figure}[t]
\centering
\includegraphics[width = 1\columnwidth, height = 3 in]{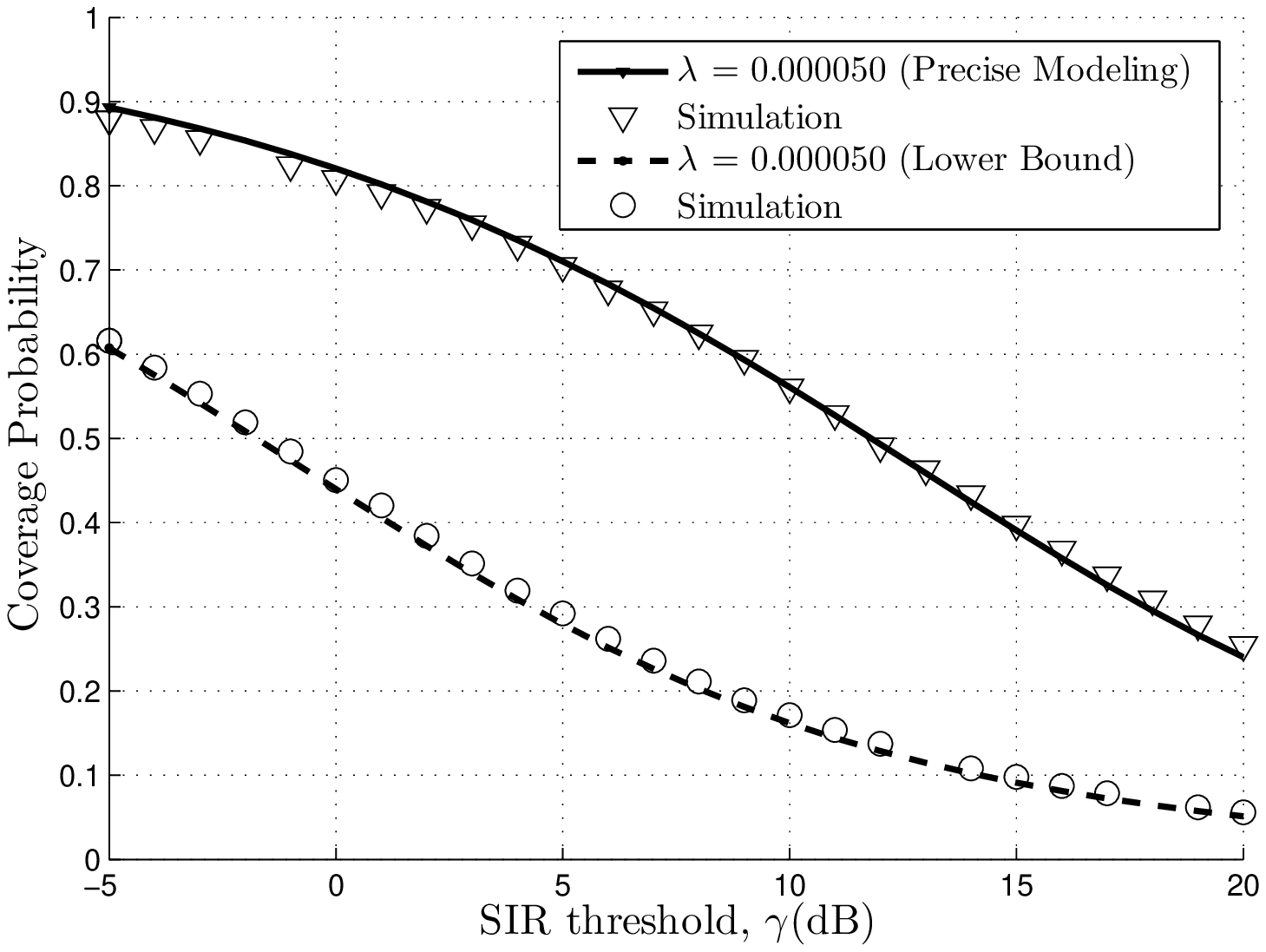}
\caption{Coverage probability of cellular user for $p_{i} = $ 0.2mW, $p_{c} = $100mW, $R_{0} =$ 1m, $r_d \leq$ 50m, and $\mu$ = 50m.}\label{Figure:theorem1}
\vspace{-1mm}
\end{figure}

The average coverage probability of cellular user in (\ref{PGFL_SC3}) depends on D2D user density $\lambda$, D2D transmit power $p_i$ and the transmit power of cellular user $p_c$. For a maximum target distance of $\mu = $ 50m, many nodes meet distance criterion, however, the pair with meets shortest distance criterion ($r_d \ll \mu$) is chosen for D2D communication. For such a scenario, it is reasonable to assume same transmit power for every D2D pair in the coverage area. To analyze the gain due to introducing retention probability, the average coverage probability of cellular user has been compared with the lower bound (conventional modeling) in Fig. \ref{Figure:theorem1}.
 In this figure, coverage gain can be observed for a cellular user. For example, for $\lambda$ = 0.00005, target SIR of $\gamma$ = -5(dB) results in average coverage gain of around 27.7\% whereas $\gamma$ = 20(dB) results in coverage gain of around 19.8\%. For higher values of $\lambda$ and lower $\gamma$, the coverage drop is insignificant unlike higher values of $\gamma$ which scales the effect of interference more significantly. For example, $\lambda$ = 0.000075 and $\gamma$ = 20(dB) results in coverage gain of only 8.7\%.

\begin{figure}[t]
\centering
\includegraphics[width = 1\columnwidth, height = 3 in]{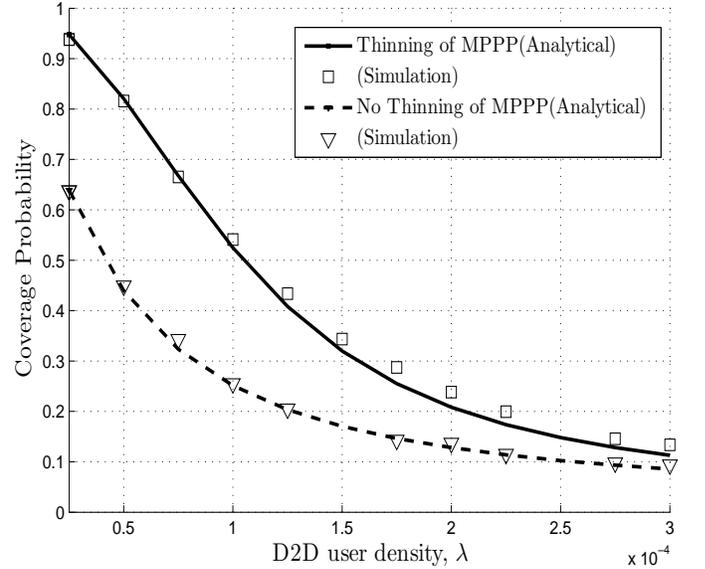}
\caption{Coverage probability of cellular user for variable D2D density $\lambda$, $r_d \leq $50m, and $\mu$ = 50m.}\label{Figure:varlambdath1}
\vspace{-3mm}
\end{figure}

In Fig. \ref{Figure:varlambdath1}, the average coverage probability of cellular user for $\gamma$ = 0(dB) and variable D2D density has been plotted. The coverage drop due to increased number of D2D pairs meets the intuition i.e., for higher value of $\lambda$, more D2D pairs can be made which means higher SN effect and reduced average coverage probability. Another effect can be seen in case of thinning where higher values of $\lambda$ results in divergence from the analytic expression as mentioned in Sec. \ref{retprob}. In case of no-thinning, every D2D pair act as an interferer resulting in close match between analytic expression and simulation setup.
\section{Conclusions}
In this paper, we introduce retention probability in Laplace functional of MPPP as the selection criterion for D2D pairing. Based on  reduced path-loss and shortest distance between D2D pairs, the average coverage probability of cellular user has been analytically and numerically evaluated. The simulation results verified the closed-form approximations for different values of $\lambda$. The D2D user density and corresponding number of D2D pairs can be selected by maintaining the average coverage probability of cellular user. A lower bound on average coverage probability of cellular user has also been introduced where no retention probability is considered and every node is assumed to be in D2D pair. This lower bound corresponds to the conventional coverage probability of a cellular user.
\bibliographystyle{IEEEtran}
\bibliography{IEEEabrv,Coverage_Gain_D2D}
\end{document}